\documentclass[submission, Phys]{SciPost}
\pdfoutput=1
\usepackage{float}
\restylefloat{table}
\usepackage{microtype}
\usepackage{xcolor}
\usepackage[most]{tcolorbox}
\usepackage[utf8]{inputenc}
\usepackage{amsmath,amssymb,amsfonts,amsthm,mathtools,mathrsfs}
\usepackage{babel}
\usepackage[sort&compress,numbers]{natbib}
\usepackage{tocloft}

\definecolor{yellow1}{HTML}{ffffcc}
\definecolor{myviolet}{HTML}{007B8B}
\definecolor{mylavender}{HTML}{DEFEFF}
\definecolor{soft}{HTML}{fff6ed}
\definecolor{sky}{HTML}{CEFFFF}
\definecolor{indigo}{HTML}{000066}
\definecolor{myblue}{HTML}{056FB1}
\definecolor{maroon}{HTML}{db0000}
\definecolor{forest}{HTML}{06961c}
\definecolor{lime}{HTML}{fcffad}
\definecolor{mygreen}{HTML}{024f16}
\definecolor{newgreen}{HTML}{aaf50a}
\definecolor{ocre}{HTML}{eb6315}

\usepackage{hyperref}
\usepackage{prettyref}
\newcommand{\pref}{\prettyref}
\newrefformat{eq}{equation (\ref{#1})}
\usepackage{bbold}
\newtcolorbox[auto counter,number within=section]{example}[2][]{ title style={right color=myblue!20, left color=myblue},
	colback=mylavender,colframe=newgreen!5,interior style={left color=newgreen!50,right color=white},fonttitle=\bfseries,breakable,enhanced jigsaw,
	title=Remark~\thetcbcounter: #2,#1}

\newtcolorbox{side}[1]{colback=lime,
	colframe=lime!20,interior style={left color=lime,right color=white},breakable,enhanced jigsaw}
\newtcolorbox{pozor}[1]{enhanced,colback=white,
	colframe=soft!20, interior style={left color=red!15,right color=white},fonttitle=\bfseries,breakable,enhanced jigsaw,
	title=#1}
\newtcolorbox[]{definition}[1][]{enhanced,borderline west={3pt}{-3pt}{cyan},colback=white,colframe =white, interior style={left color = ocre!5, right color =white}, breakable,enhanced jigsaw}

\renewcommand{\d}{\mathrm{d}}
\hypersetup{
	colorlinks=true,
	linkcolor=ocre,
	filecolor=cyan,      
	citecolor=ocre,
    urlcolor=myblue
}
\usepackage{tensor}
\usepackage{comment}
\usepackage{parskip}
\usepackage{cleveref}

\begin{document}
\begin{samepage}
		\begin{flushleft}\huge{\textbf{Monopoles, Clarified}}\end{flushleft}
		\vspace{20pt}
		{\color{myviolet}\hrule height 1mm}
		\vspace*{10pt}
		\begin{flushleft}
		\large 	\textbf{Aviral Aggarwal}$\,{}^a$, \textbf{Subhroneel Chakrabarti}$\,{}^a$, \textbf{and Madhusudhan Raman}$\,{}^b$
		\end{flushleft}

		\begin{flushleft}
			\emph{\large ${}^a$ Department of Theoretical Physics and Astrophysics,\\ Faculty of Science, Masaryk University, 611 37 Brno, Czech Republic.}
			\\ \vspace{1mm}
            \large ${}^b$ \emph{Department of Physics and Astrophysics\\
            University of Delhi, Delhi 110 007, India}
			\\ \vspace{3mm}
             \href{mailto:aviral@mail.muni.cz}{aviral@mail.muni.cz}, \href{mailto:subhroneelc@physics.muni.cz}{subhroneelc@physics.muni.cz},  \href{mailto:mraman@physics.du.ac.in}{mraman@physics.du.ac.in} \\
		\end{flushleft}
		
  \begin{flushright}
			%\emph{Date: \today}
		\end{flushright}
		\section*{Abstract}
		{\bf
			We propose a manifestly duality-invariant, Lorentz-invariant, and local action to describe quantum electrodynamics 
            in the presence of magnetic monopoles that derives from Sen's formalism. By employing field strengths as the dynamical variables, rather than potentials, this formalism resolves longstanding ambiguities in prior frameworks. Our analysis finds consistent outcomes at both tree and loop levels using the established principles of quantum field theory, obviating the need for external assumptions or amendments. We clarify the mechanisms of charge renormalisation and demonstrate the renormalisation group invariance of the charge quantisation condition. Our approach can be useful for phenomenological studies and in quantum field theories with strong-weak dualities.
		}
\end{samepage}
	\newpage
	\vspace{10pt}
	\noindent\rule{\textwidth}{1pt}
	\pagecolor{white}
	\tableofcontents\thispagestyle{fancy}
	\noindent\rule{\textwidth}{1pt}
	\vspace{10pt}
\section{Introduction} \label{sec:intro}

Magnetic monopoles have been a fecund source of insight and
inspiration to theoretical physicists and mathematicians since Dirac's
pioneering investigations into the quantum theory of magnetic
monopoles \cite{Dirac:1931kp}. The subsequent discovery of topological
monopole solutions in spontaneously broken gauge theories by 't Hooft
\cite{tHooft:1974kcl} and Polyakov \cite{Polyakov:1974ek} and the
implications of their existence for grand unified theories and
cosmology (see, for example, \cite{Zeldovich:1978wj,Preskill:1979zi}) are just some of
the reasons why monopoles, despite evading detection for decades
\cite{Milton:2006cp}, remain an active subject of study.

Quantum field theories with interacting electric and magnetic charges
are subtle for reasons that are useful to recall, if only
briefly. (For a more comprehensive review, we refer the reader to
\cite{Blagojevic:1985sh}.) This story begins with Dirac's observation
that the source-free Maxwell equations:
\begin{equation}
\partial _{\mu }F ^{\mu \nu } = 0 \quad \mathrm{and} \quad \partial _{\mu }\widetilde{F} ^{\mu \nu } = 0 \ ,
\end{equation}
where
$ \widetilde{F} _{\mu \nu } = \frac{1}{2} \epsilon _{\mu \nu \rho
  \sigma }F ^{\rho \sigma } $ is the Hodge dual of the electromagnetic
field strength, are invariant under the exchange
\begin{equation}
F ^{\mu \nu } \leftrightarrow \widetilde{F} ^{\mu \nu } \ .
\end{equation}
In order to extend this invariance to Maxwell theory in the presence
of sources, one \emph{must} posit the existence of magnetic
monopoles. Further, as demonstrated by Dirac \cite{Dirac:1931kp}, any
quantum theory of electric ($ e $) and magnetic ($ g $) charges must obey charge
quantisation: the product of electric and magnetic charges must satisfy:
\begin{equation}
e g = 2 \pi n \quad \mathrm{for} \quad n \in \mathbb{Z} \ .
\end{equation}
Indeed, as pointed out by Saha \citep{Saha1936}, based on the classical derivation by Thomson in 1904 \citep{Thomson1904}, charge quantisation can also be equivalently derived by considering a semiclassical quantisation of the total angular momentum of a charge-monopole system. This derivation has the advantage of not referring to gauge potential in deriving the result, which will be particularly relevant for our approach. Since (electric and magnetic) charges set the coupling strength
between light and charged matter, this elegant result implies a
nonperturbative relation between a strongly coupled theory with a
weakly coupled electric-magnetic dual.

It is therefore of interest to study theories with dynamical electric
and magnetic sources --- sometimes referred to as quantum
electro-magnetodynamics (QEMD) --- in a way that preserves manifest Lorentz
\emph{and} duality invariance. (Such theories are known to exist, for
example, at Argyres-Douglas points in the moduli space of
$ \mathcal{N}=2 $ supersymmetric gauge theories
\cite{Argyres:1995jj}.) Satisfying these twin requirements is not
easy, and it is easy to understand why: while duality acts covariantly
on field strengths, it does not act on gauge potentials --- in terms
of which Maxwell theory is typically formulated --- in a local,
Lorentz-covariant manner. Over the past few decades, many attempts
have been made to circumvent this difficulty
\cite{Dirac:1948um,Zwanziger:1968rs,Zwanziger:1970hk,Deser:1976iy,Schwarz:1993vs,Pasti:1995ii,Mkrtchyan:2019opf};
a feature often shared by these proposed resolutions is the sacrifice
of either manifest Lorentz invariance or locality, or having to work with an action that is not straightforwardly quantisable. 

Similar difficulties plague attempts to find local, Lorentz invariant actions for self-dual $ (2n+1) $-form field-strengths in $ (4n+2) $-dimensions. Indeed, it was soon realised that these two classes of field theories suffering from identical ailments are intimately related \citep{Schwarz:1993vs,Berkovits:1996nq,Berkovits:1996rt,Berkovits:1996em,Lechner:1999ga,Avetisyan:2021heg,Avetisyan:2022zza}. On compactifying the quantum field theory of a self-dual field strength in $(4n+2)$ dimensions on a torus, one obtains an electromagnetic duality invariant theory in $(4n)$-dimensions. The past decades have seen a slew of attempts to find suitable actions for self-dual field strengths; all but one fail to give rise to a local, manifestly Lorentz and duality invariant, polynomial, and straightforwardly quantisable action. The only formalism that passes all these checks is due to Sen \cite{Sen:2015nph,Sen:2019qit} and has come to be known as Sen's formalism. Sen's formalism has been used in a wide variety of theories, in various dimensions, with and without gravity \cite{Lambert:2019diy,Andriolo:2020ykk,Gustavsson:2020ugb,Vanichchapongjaroen:2020wza,Rist:2020uaa,Chakrabarti:2020dhv,Andriolo:2021gen,Andrianopoli:2022bzr,Barbagallo:2022kbt,Chakrabarti:2022lnn,Chakrabarti:2022jcb,Chakrabarti:2023czz,Lambert:2023qgs,Phonchantuek:2023iao,Hull:2023dgp}.\footnote{See \citep{Evnin:2022kqn} for a comparison of some of the approaches including Sen's formalism.}

It is therefore natural to wonder if Sen's formalism can be leveraged to obtain a suitable action for the case of QEMD. Indeed, in \citep{Sen:2019qit} such an action was obtained on a general four-dimensional Lorentzian curved manifold by compactifying a self-dual $3$-form theory on a $2$-torus. However, the analysis in \citep{Sen:2019qit} restricted itself to just writing down the action and the Hamiltonian and noting its desirable qualities. Moreover, the action presented there requires considerable understanding of the parent six-dimensional theory. In this article, we propose a rewriting of Sen's action that can be presented in a self-contained way in four dimensions itself; given that it has all the desirable features, we proceed from this starting point to ask and definitively answer a number of questions that have been raised in the recent literature on the subject using \textit{nothing} but textbook perturbative quantum field theory techniques.

Sen's formalism has a couple of unusual features from the perspective of standard approaches to Abelian gauge theories. First, maintaining manifest Lorentz symmetry in the coordinate space action requires that we introduce additional, dynamical fields. But the structure of the theory is such that these additional fields do not couple to the physical fields of interest (i.e.~electric and magnetic fields and their sources). This decoupling is guaranteed at the level of the full quantum Hamiltonian. But it is much simpler to see at the level of equations of motion and the momentum space action (which is what we use to derive the Feynman rules). Second, Sen's formalism treats the field strength $F_{\mu \nu}$ directly as a dynamical variable, and we do not invoke the notion of gauge potential at any point of the calculation. At first glance, while this is unusual, a little thought reveals that indeed this is a welcome feature to have for QEMD. It has long been understood that the fact that Bianchi identities no longer hold in QEMD implies that gauge potentials, as usually thought of in QED, are no longer good dynamical variables. This is, of course, not a new idea, see
\cite{Halpern:1978ik,Calucci:1981we,Calucci:1982fm,Calucci:1982wy,Blagojevic:1985yd}
for earlier applications of this formalism in position space. 

Recently, \citep{Newey:2024gug} the authors approached the task of understanding how magnetic charges are renormalised; however, they used an approach that is not manifestly local and Lorentz invariant. Even so, the authors of \citep{Newey:2024gug} astutely noted that many of the apparently puzzling observations that seem to stem from lack of locality and Lorentz symmetry can be avoided if one reformulates the Feynman rules in terms of field strengths themselves. In our analysis, this is in-built and natural, and we will at all times maintain locality, Lorentz invariance, and duality symmetry. Keeping all the symmetries manifest will reveal that there never was any puzzle or paradox in QEMD.

The outline of the rest of this paper is as follows. We write down in \Cref{sec:action-symmetries} a manifestly local, Lorentz- and
duality-invariant action for QEMD that reproduces the extended Maxwell
equations in four-dimensional Minkowski spacetime. We also discuss all the (gauge) symmetries that the action possesses and show the decoupling of the extra fields at the level of equations of motion. In \Cref{sec:feynman-rules}, we derive the Feynman rules systematically for
subsequent use in perturbation theory. We establish that the momentum space action has the extra fields completely decoupled, thereby allowing us to essentially forget about them in any scattering amplitude computation involving the physical fields and currents. We then go on to compute all tree-level current-current scattering amplitudes, thereby reproducing the classic action-independent results of Weinberg
\cite{Weinberg:1965rz} and show how the apparent violation of Lorentz (and indeed gauge) symmetry in electric-magnetic scattering is automatically resolved without any additional inputs. Given that we have a simple action for QEMD that correctly reproduces tree-level propagators and current correlation functions, it is
natural to ask whether we can incorporate quantum mechanical corrections. This task is complicated by charge quantisation, a point that is well-appreciated in the literature, even if this appreciation has not been accompanied by clarity or consensus on how electric and magnetic charges are renormalised
\cite{Schwinger:1966zza,Schwinger:1966zzb,Brandt:1977fa,Deans:1981qs,Panagiotakopoulos:1982fp,Coleman:1982cx,Calucci:1982wy,Jengo:1982wx,Goebel:1983we,Newey:2024gug}. We
offer a clean derivation in \Cref{sec:charge-renormalisation} of the renormalisation group invariance of the charge quantization rule based only on standard perturbative techniques. Additionally, we provide a critical appraisal of previous attempts to answer this question and provide a definitive resolution to this long-standing conundrum. We conclude with a short summary of our key results and a few exciting future problems that we wish to return to in the near future. Finally, Appendix \pref{app:Sen} explains how duality symmetry appears at the level of Maxwell's equations when the sources are expressed in terms of $2$-forms.

\section{The Action and Its Symmetries}
\label{sec:action-symmetries}
\subsection{Sen's Formalism and Dimensional Reduction}

Let us present a summary of the results of \citep{Sen:2019qit} where an action of self-dual $3$-form in six dimensions was dimensionally reduced over a torus to obtain an electric-magnetic duality invariant action in four dimensions. The purpose of this section is twofold. First, we hope this summary of Sen's action and its nuances will help this article be more self-contained. Second, it will also help set up our notation and conventions, which differ slightly from \citep{Sen:2019qit}; in particular, we express the sources in a way that is more natural from the four-dimensional perspective.

Sen's action in six dimensions describes the dynamics of a self-dual $3$-form $H_{abc}$, where the indices $\{a,b,c,\cdots \}$ run over $\{0,1,\cdots,5\}$. The two essential novelties in Sen's formalism are

\begin{itemize}
    \item[(i)] The field $H_{abc}$ is itself treated as the dynamical variable and it is not thought of as a \textit{field strength} of a $2$-form gauge potential.
    \item[(ii)] To write down a manifestly Lorentz invariant kinetic term for the field $H_{abc}$ one needs to introduce an additional dynamical 2-form field $P_{ab}$ that only enters the action via its field strength $\partial_{[a}P_{bc]}$.
\end{itemize}

However, as was proven in \citep{Sen:2019qit}, these extra fields completely decouple from the dynamics of the self-dual field, even in presence of arbitrary interactions (as long as the interaction terms are independent of the extra fields) at the level of the full quantum Hamiltonian. Therefore, Sen's action presents a manifestly Lorentz invariant and easily quantisable action for interacting self-dual fields.

For the purposes of this article it is sufficient to focus on Sen's action in flat spacetime and the case where the only interactions are due to classical sources for the field $H_{abc}$. Note that, unlike the standard textbook formulations, the sources for this field-strength-like variable cannot be the usual currents, but rather a $3$-form itself, which we will denote $\Omega_{abc}$.

The action in six dimensions reads as follows:

\begin{equation}
    S_{6\mathrm{d}} = \int \mathrm{d}^{6}x \bigg[\frac{1}{3!} \partial_{[a}P_{bc]}\partial^{[a}P^{bc]}  - \partial_{[a}P_{bc]} H^{abc} - H_{abc} \Omega^{abc}  \bigg] \quad.
\end{equation}

It is important to notice that the kinetic terms for the extra fields come with the wrong sign. (We work in the mostly-plus metric
convention.) This wrong sign is crucial for the decoupling of these fields. 

We now consider the dimensional reduction of the above action on a $2$-torus, which we take to be along the directions $\{x^4,x^5\}$. Furthermore, we freeze the dynamics on these two directions, so no fields or sources depends on the torus directions. It was shown explicitly in \citep{Sen:2019qit} that a consistent truncation of the theory is obtained by only looking at the following fields: $P^A_{\mu}$, $H^{A}_{\mu \nu}$, and $\Omega^{A}_{\mu \nu}$. Here, we have denoted the $\{x^4,x^5\}$ directions by the index $A$, and the remaining four-dimensional spacetime indices by the Greek alphabets $\{\mu,\nu,\rho,\cdots \}$.

Therefore on four-dimensional spacetime, the field content is as follows.

\begin{itemize}
    \item Two $1$-form extra fields coming from $P^A_\mu$ which we denote $ B ^{(1)}_\mu $ and $ B ^{(2)}_\mu $. 

    \item Naively, it would seem like we get two $2$-form fields from $H^A_{\mu \nu}$; recall, however, that $H_{abc}$ is self-dual. This implies that the four-dimensional avatars are $ 2 $-form field strength $ F _{\mu \nu } $ and its four-dimensional Hodge dual
$ \widetilde{F}_{\mu \nu } = \frac{1}{2} \epsilon _{\mu \nu \rho  \sigma }F ^{\rho \sigma }$.

    \item Finally, the source $\Omega^{A}_{\mu \nu}$ being an unconstrained $3$-form would give rise to a pair of $2$-form fields. However, note that in the action $\Omega$ enters via inner product with $H$, which is self-dual. Therefore, only the anti-self-dual part of $\Omega$ enters the action. Thus once again we get on dimensional reduction a $2$-form source $\Sigma_{\mu \nu}$ and its Hodge dual $\widetilde{\Sigma}_{\mu \nu}$.
\end{itemize}

It can be easily checked that Sen's action, on dimensional reduction, and with the above-mentioned truncation reduces to the following four-dimensional action (after minor numerical rescaling):
\begin{equation}
  \label{eq:our-action}
\begin{aligned}
  S = \int _{} ^{} \mathrm{d}^{4}x \, &\bigg[ \frac{1}{4} G ^{(1)}_{\mu \nu }G ^{(1)\mu \nu } + \frac{1}{4} G ^{(2)}_{\mu \nu }G ^{(2)\mu \nu } - G ^{(1)}_{\mu \nu } F ^{\mu \nu } - G ^{(2)}_{\mu \nu } \widetilde{F} ^{\mu \nu }  
  - \widetilde{F}_{\mu \nu } \Sigma ^{\mu \nu } -  F _{\mu \nu } \widetilde{\Sigma }^{\mu \nu } \bigg] \ .
\end{aligned}
\end{equation}
where we have defined
\begin{equation}
G ^{(i)}_{\mu \nu } = \partial _{\mu }B ^{(i)}_{\nu } - \partial _{\nu }B ^{(i)}_{\mu } \quad \mathrm{for} \quad i = 1,2\ .
\end{equation}

\subsection{Equations of Motion, Gauge Symmetries, and Duality Invariance}
The equations of motion that arise from this action are
\begin{equation}
  \label{eq:eoms-B-F}
\begin{aligned}
  \delta _{B _{1}}S = 0 \quad &\Rightarrow \quad \partial _{\mu }G ^{(1)\mu \nu } = 2\partial _{\mu }F ^{\mu \nu } \ , \\
  \delta _{B _{2}}S = 0 \quad &\Rightarrow \quad \partial _{\mu }G ^{(2)\mu \nu } = 2 \partial _{\mu } \widetilde{F}^{\mu \nu } \ , \\
  \delta _{F} S = 0 \quad &\Rightarrow \quad  G ^{(1)}_{\mu \nu } + \widetilde{G} ^{(2)}_{\mu \nu } + 2 \widetilde{\Sigma }_{\mu \nu } = 0 \ .
\end{aligned}
\end{equation}
The last of these equations also implies, via Hodge duality:
\begin{equation}
  \label{eq:eoms-FStar}
 - \widetilde{G} ^{(1)}_{\mu \nu } + G ^{(2)}_{\mu \nu } + 2 \Sigma _{\mu \nu } = 0\ ,
\end{equation}
where we have used the fact that the Hodge dual $\star$ satisfies $ \star ^{2} = -1 $.  

Using the above \cref{eq:eoms-B-F,eq:eoms-FStar} to eliminate $ B _{1}
$ and $ B _{2} $, we get the equations of motion
\begin{equation}
  \label{eq:extended-Maxwell}
\begin{aligned}
  \partial _{\mu }F ^{\mu \nu } &= - \partial _{\mu }\widetilde{\Sigma } ^{\mu \nu } \ , \\
  \partial _{\mu } \widetilde{F} ^{\mu \nu } &= -\partial _{\mu } \Sigma ^{\mu \nu } \ ,
\end{aligned}
\end{equation}
which are the extended Maxwell equations in the presence of electric
and magnetic sources. The ``traditional'' form of the extended Maxwell
equations is obtained under the identifications
\begin{equation} \label{eq:sigma_J}
\begin{aligned}
  \partial _{\mu } \widetilde{\Sigma } ^{\mu \nu } &= - e J _{e}^{\nu } \ , \\
  \partial _{\mu } \Sigma ^{\mu \nu } &= - g J _{m}^{\nu } \ ,
\end{aligned}
\end{equation}
where $ J _{e} $ and $ J _{m} $ are the electric and magnetic currents
that would couple in the usual way if one were working in the
potential formalism and $e$ and $g$ are the electric and magnetic coupling constants respectively. Note that, in flat spacetime, any conserved vector $V_\mu$ can be expressed as a $4$-divergence of an antisymmetric tensor. We thus identify this as the relation between our source $\Sigma_{\mu\nu}$ and its dual $\widetilde{\Sigma}_{\mu\nu}$ to the electric and the magnetic current.

The observant reader might be bothered by the fact that the electric and magnetic $2$-form sources are related by Hodge duality, whereas the usual currents $J_e$ and $J_m$ might not be a priori related. This is easily explained. First of all, as we mentioned earlier, this is inevitable in Sen's formalism since only the anti-self-dual part of the source enters the action. Additionally, this fact can also be demonstrated directly from the extended Maxwell equations. The proof does not have any bearing on our subsequent analysis, but for completeness, we have presented the details in Appendix \pref{app:Sen}.

Finally, if we define $H^{(1)}_{\mu \nu} := G^{(1)}_{\mu \nu} - 2 F_{\mu \nu}$ and $H^{(2)}_{\mu \nu} := G^{(2)}_{\mu \nu} - 2 \widetilde{F}_{\mu \nu}$, then these new fields can now be interpreted as encoding the additional degrees of freedom, and from \pref{eq:eoms-B-F} it follows that they satisfy
\begin{equation}
    \partial^\mu H^{(1)}_{\mu \nu} = 0 \;,\; \partial^\mu H^{(2)}_{\mu \nu} = 0 
\end{equation}
establishing that they are a decoupled pair of free fields that do not interact with the physical sector, at least at the level of classical solutions. Of course, as we mentioned earlier, these extra degrees of freedom also decouple at the level of full quantum Hamiltonian. We will see later in \pref{sec:feynman-rules} that this decoupling is also easily seen at the level of the momentum space action where we derive the Feynman rules.

Our action has two types of gauge symmetries. The first one is the usual $\mathrm{U}(1)$ gauge symmetries of the pair of extra gauge fields, viz.
\begin{equation}
    \delta_\phi B^{(i)}_\mu = \partial_\mu \phi^{(i)} \quad \mathrm{for} \quad i =1,2.
\end{equation}

The action is also invariant (up to irrelevant boundary terms) under a different set of gauge transformations:
\begin{equation}
\begin{aligned}
\delta_{v} B^{(1)}_{\mu }=v ^{(2)}_{\mu } \quad &\mathrm{and} \quad \delta_{v} B ^{(2)}_{\mu } = -v ^{(1)}_{\mu } \ , \\
\delta F _{\mu \nu } = \partial _{[\mu} v ^{(2)}_{ \nu ]}&+ \frac{1}{2} \epsilon _{\mu \nu \rho \sigma } \partial ^{\rho } v ^{(1)\sigma } \, , \\
2 \delta \Sigma _{\mu \nu } = \partial _{[\mu }  v ^{(1)}_{\nu ]} &+\frac{1}{2} \epsilon _{\mu \nu \rho \sigma } \partial ^{\rho } v ^{(2)\sigma }\ , 
\end{aligned}
\end{equation}
with the $1$-form gauge parameters subject to the conditions
\begin{equation}
  \begin{aligned}
    \int \mathrm{d}^{4}x \, \epsilon ^{\mu \nu \rho \sigma } v ^{(2)}_{\mu } \partial _{\nu } \Sigma _{\rho \sigma } = 0 \ , \\
    \int \mathrm{d}^{4}x \, \epsilon ^{\mu \nu \rho \sigma } v ^{(1)}_{\mu } \partial _{\nu } \widetilde{\Sigma } _{\rho \sigma } = 0 \ .
  \end{aligned}
\end{equation}
These conditions on $1$-form gauge parameters $v ^{(1)}$ and
$ v ^{(2)} $ ensure that the additional fields $B^{(1)}$ and $ B ^{(2)} $
are not pure gauge.

Typically in the literature, actions written in Sen's formalism keep the coupling constants implicit in the sources. However, since we wish to do perturbative quantum field theory with this action, it will be helpful to restore the coupling constants explicitly. This is readily suggested from \pref{eq:sigma_J}. We define 

\begin{align}
    \widetilde{\Sigma}_{\mu \nu} &= -e \Sigma^e_{\mu \nu} \nonumber \\
    \Sigma_{\mu \nu} &= - g \Sigma^m_{\mu \nu} \quad.
\end{align}

Note that the electric and magnetic sources thus defined are related now by the Hodge duality up to a proportionality factor, viz.

\begin{equation} \label{eq:source_duality}
    g\widetilde{\Sigma}^m_{\mu \nu} = - e \Sigma^e_{\mu \nu} \quad, \quad e\widetilde{\Sigma}^e_{\mu \nu} = g\Sigma^m_{\mu \nu} \;.
\end{equation}

With these sources, the action now takes the form

\begin{equation}
  \label{eq:our-action_new}
  S = \int _{} ^{} \mathrm{d}^{4}x \, \bigg[ \frac{1}{4} G ^{(1)}_{\mu \nu }G ^{(1)\mu \nu } + \frac{1}{4} G ^{(2)}_{\mu \nu }G ^{(2)\mu \nu } -  G ^{(1)}_{\mu \nu } F ^{\mu \nu } -  G ^{(2)}_{\mu \nu } \widetilde{F} ^{\mu \nu }  
  + g \widetilde{F}_{\mu \nu } \Sigma_m ^{\mu \nu } + e F _{\mu \nu } \Sigma_e^{\mu \nu } \bigg] \ .
\end{equation}

Under an electric-magnetic duality operator $ \mathcal{S} $, which
acts as a homomorphism, any quantity $ \bullet $ in our theory
transforms as $ \bullet \rightarrow \mathcal{S}\left[\bullet\right] $
with
\begin{equation}
\begin{aligned}
  \mathcal{S}\left[F ^{\mu \nu }\right] &= - \widetilde{F}^{\mu \nu } \ , \\
  \mathcal{S}\left[e\Sigma_e ^{\mu \nu }\right] &= -g\Sigma_m ^{\mu \nu } \\
\end{aligned}
\end{equation}
Together with the property that electric-magnetic $ (\mathcal{S}) $
and Hodge $ (\star ) $ duality commute,
\begin{equation}
\star \, \mathcal{S}\left[ \bullet \right] = \mathcal{S}\left[ \star \, \bullet \right] \ ,
\end{equation}
we find that the extended Maxwell equations in
\cref{eq:extended-Maxwell} are invariant, as they should be. At the
level of the action, one needs to specify a duality transformation for
the extra fields that is consistent with the aforementioned
duality properties, and it is easy to see that the following assignment:
\begin{equation}
\begin{aligned}
  \mathcal{S}\left[B ^{(1)}_{\mu }\right] &= - B ^{(2)}_{\mu } \ , \\
  \mathcal{S}\left[B ^{(2)}_{\mu }\right] &= B ^{(1)}_{\mu } \ ,
\end{aligned}
\end{equation}
is sufficient to leave the action duality invariant. This form of the duality transformation of the extra fields is also the one that is suggested from the parent six-dimensional theory, where the duality transformation is nothing but an exchange of the two torus directions. In fact our action, as written in \cref{eq:our-action_new}, can be expressed as sum of two actions as follows

\begin{align} \label{eq:S_e_action}
    S &= S_e + S_m  \;,\nonumber \\
    S_e &=  \int _{} ^{} \mathrm{d}^{4}x \, \bigg[ \frac{1}{4} G ^{(1)}_{\mu \nu }G ^{(1)\mu \nu } -  G ^{(1)}_{\mu \nu } F ^{\mu \nu }  + e F _{\mu \nu } \Sigma_e^{\mu \nu } \bigg] \ , \nonumber \\
    S_m &= \int _{} ^{} \mathrm{d}^{4}x \, \bigg[\frac{1}{4} G ^{(2)}_{\mu \nu }G ^{(2)\mu \nu } -  G ^{(2)}_{\mu \nu } \widetilde{F} ^{\mu \nu }   + g \widetilde{F}_{\mu \nu } \Sigma_m ^{\mu \nu }  \bigg] \ \ .
\end{align}

It is easy to see that under duality $\mathcal{S}[S_e] = S_m$ and $\mathcal{S}[S_m] = S_e$, leaving the total action duality invariant.

If one wishes to make duality property even more manifest, it is best to introduce the following doublets
\begin{align}
    \mathcal{G} &= \begin{pmatrix}
        G^{(1)} \\ -G^{(2)}
    \end{pmatrix} \, ,\\
    \mathcal{F} &= \begin{pmatrix}
        F \\ -\widetilde{F}
    \end{pmatrix} \; , \\
    \widehat{\Sigma} &= \begin{pmatrix}
        e\Sigma_e \\ -g\Sigma_m 
    \end{pmatrix}    \;.
\end{align}

We have momentarily suppressed the Lorentz indices to avoid clutter. Notice that the action of the duality operator $\mathcal{S}$ on the fields and the source is now captured by the $2 \times 2$ matrix

\begin{equation}
    \mathcal{S} = \begin{pmatrix}
        0 & 1 \\ -1 & 0
    \end{pmatrix} \;.
\end{equation}

The action in \pref{eq:our-action_new} now becomes in terms of these doublets (full contraction of Lorentz indices implicitly assumed)

\begin{equation}
    S = \int \mathrm{d}^{4} x\bigg[ \frac{1}{4} \mathcal{G}^T \mathcal{G}  - \mathcal{G}^T \mathcal{F} - \mathcal{F}^T\widehat{\Sigma}                                     \bigg] \;.
\end{equation}
Using $\mathcal{S}^T   \mathcal{S} = \mathbb{1}$ it is easy to check the action is now manifestly duality invariant. In fact, the \emph{full} $\mathrm{SO}(2)$ electric-magnetic duality group is manifest.

\section{The Feynman Rules}
\label{sec:feynman-rules}
A local, duality- and Lorentz-invariant action should lend itself to a straightforward derivation of Feynman rules, which may be used to perform perturbative computations without any additional inputs. The goal of this section, therefore, is twofold. First, we derive the Feynman rules from the action introduced in \Cref{sec:action-symmetries}. Then we proceed to use those Feynman rules to reproduce the tree-level current-current correlation functions and show that they match with the Lagrangian-independent computation presented in \citep{Weinberg:1965rz}. We will see that the manifest duality symmetry of our formalism makes the confusion regarding violation of Lorentz and gauge symmetry for electric-magnetic scattering, dubbed the ``Weinberg Paradox'' \citep{Terning:2018udc}, disappear entirely. 

Let us first focus on the propagator and therefore temporarily restrict our attention to the kinetic terms in \cref{eq:our-action}:
\begin{equation}
    S \supseteq \frac1{4} \int d^4x \,  G^{(1)}_{\mu \nu} \, G^{(1)\mu \nu}+ \frac1{4} \int d^4x \,G^{(2)}_{\mu \nu}\,G^{(2)\mu \nu} - \int d^4x \, G^{(1)}_{\mu \nu}\,  F^{\mu \nu}- \int d^4x \, G^{(2)}_{\mu \nu} \, \tilde{F}^{\mu \nu} \;.
\end{equation}
As in any gauge theory, we need to gauge fix before we invert the kinetic term to find the propagator. For our action, it is enough to gauge fix the two $\mathrm{U}(1)$ gauge invariances associated with the fields $B^{(1)}$ and $B^{(1)}$. We do this by introducing the standard Lorenz gauge fixing term
\begin{equation}
    S_{\mathrm{gf}}=\sum_{i=1,2}\frac{1}{2}\int \big(\partial^{\mu}B^{(i)}_{\mu}\big)^2 \;.
\end{equation}
The gauge fixed action therefore assumes the form 
\begin{equation}
   S\supseteq\sum_{i=1,2} \int \mathrm{d}^4x \, \left[ \frac{1}{2} \partial_{\mu} B^{(i)}_{\nu}\partial^{\mu} B^{(i)\nu}- 2 B^{(1)}_{\nu} \partial_{\mu} F^{\mu \nu}- 2 B^{(2)}_{\nu} \partial_{\mu} \widetilde{F}^{\mu \nu}\right] \ .
\end{equation}

While in coordinate space there is a kinetic mixing between the physical photon fields and the additional gauge fields $B^{(i)}$, we will now show that in the momentum space there is a simple field redefinition that completely decouples the kinetic terms for the photons and the additional fields. First, we rewrite the gauge-fixed action in momentum space:\footnote{Our conventions for Fourier transforms are those implied by the replacement $i\partial \rightarrow k $.}
\begin{align}
    S &\supseteq\sum_{j=1,2} \frac{1}{2} \int \frac{\mathrm{d}^4k}{(2 \pi)^4} B^{(j)}_{\mu}(k) k^2 B^{(j)\mu}(-k)+2 i \int \frac{\mathrm{d}^4k}{(2 \pi)^4}  F^{\mu \nu }(-k) k_{\mu} B^{(1)}_{\nu}(k) \nonumber \\
    &\quad+ 2 i \int \frac{\mathrm{d}^4k}{(2 \pi)^4} \widetilde{F}^{\mu \nu }(-k) k_{\mu} B^{(2)}_{\nu}(k) \;. 
\end{align} 
Let us now define the following fields
\begin{equation}
    \begin{aligned}
        &\widehat{B}^{(1)}_{\mu}(k) := B^{(1)}_{\mu}(k) -\frac{2i }{k^{2}}\,k^{\rho}F_{\rho \mu}(k) \ , \\
        &
       \widehat{B}^{(2)}_{\mu}(k) :=  B_{\mu}^{(2)}(k) -\frac{2i }{k^{2}} \,k^{\rho}\widetilde{F}_{\rho \mu}(k) \ , 
    \end{aligned}
\end{equation}

The gauge-fixed action in terms of this new fields now becomes
\begin{align}\label{eq:kinetic_term}
     S  &\supseteq S_B + S_F \nonumber \ , \\
     &= \sum_{i=1,2} \frac{1}{2}  \int \frac{\mathrm{d}^4k}{(2 \pi)^4} \widehat{B}^{(i)\mu}(k) k^2\widehat{B}^{(i)}_{\mu}(-k) \nonumber\\
     &\quad +\int \frac{\mathrm{d}^4k}{(2 \pi)^4}\frac{2 k^{\rho} F_{\rho \mu }(k)k_{\nu} F^{\nu \mu }(-k) }{k^2}+ \int \frac{\mathrm{d}^4k}{(2 \pi)^4} \frac{2 k ^{\rho} \widetilde{F}_{\rho \mu }(k)k_{\nu} \widetilde{F}^{\nu \mu }(-k)}{k^2} \; .
\end{align}
We see that the action has now decoupled into two decoupled kinetic terms, one for the additional fields $\widehat{B}^{(i)}$s and the other corresponding to the physical photon field.
Recall further that the source terms in \cref{eq:our-action} only coupled to the photon, so there will be no interaction vertices in our action that involve the additional fields. Therefore, we can henceforth safely forget about the additional fields since they will not be involved in any physical scattering processes involving photons and currents. 

The dynamics of the photon field is captured by both $F_{\mu \nu}$ and its dual $\widetilde{F}_{\mu \nu}$. However, due to charge quantisation, it is only allowed to perturbatively expand about only one of the couplings, either $e$ or $g$. If we expand in $e$, then the vertices emit $F$ and we should use the $F$-$F$ propagator. On the other hand, if the magnetic charge $g$, is perturbatively coupled, the vertex emits an $\widetilde{F}$ and we should use the $\widetilde{F}$-$\widetilde{F}$ propagator. Of course, the two propagators are identical as functions of momentum as they should be due to the duality symmetry. The duality symmetry of the theory ensures that observables in either choice can be related to the other.

Let us focus on the kinetic term for $F_{\mu \nu}$ for specificity. It can be expressed as
\begin{equation}
    S= \frac{1}{2}\int \frac{\mathrm{d}^4k}{(2 \pi)^4}  F^{\mu \nu}(k) \, \mathcal{K}_{\mu\nu}\,^{\rho\sigma}(k) \, F_{\rho \sigma}(-k) \;.
\end{equation} 
where we have defined
\begin{equation}
    \mathcal{K}_{\mu\nu}\,^{\rho\sigma}(k):=  4\frac{k_{[\mu}k^{[\rho}\delta_{\nu]}\,^{\sigma]}}{k^2}\;.
\end{equation}

The propagator is now easily read off from the kinetic operator (throughout we will suppress the Feynman $i\epsilon$ factor for readability):
\begin{equation}
    i\Delta_{\mu \nu}^{\;\;\;\;\rho \sigma} = \left\langle F_{\mu \nu} F^{\rho \sigma}\right\rangle =  2i \frac{k_{[\mu}k^{[\rho}\delta_{\nu]}\,^{\sigma]}}{k^2} \;.
\end{equation}
The vertex factors are easily obtained from \cref{eq:our-action_new}. We have two kinds of vertices: A vertex for the coupling of the photon to the electric source
\begin{equation} \label{eq:vertex_source}
    V^e_{\mu \nu}=  i e \Sigma^e_{\mu \nu} \ ,
\end{equation}
and a vertex for the coupling to the magnetic source 
\begin{equation}
    V^m_{\mu \nu} = ig \Sigma^m_{\mu \nu} \;.
\end{equation}

\begin{figure}[H]
    \centering
    \includegraphics[width=0.5\textwidth]{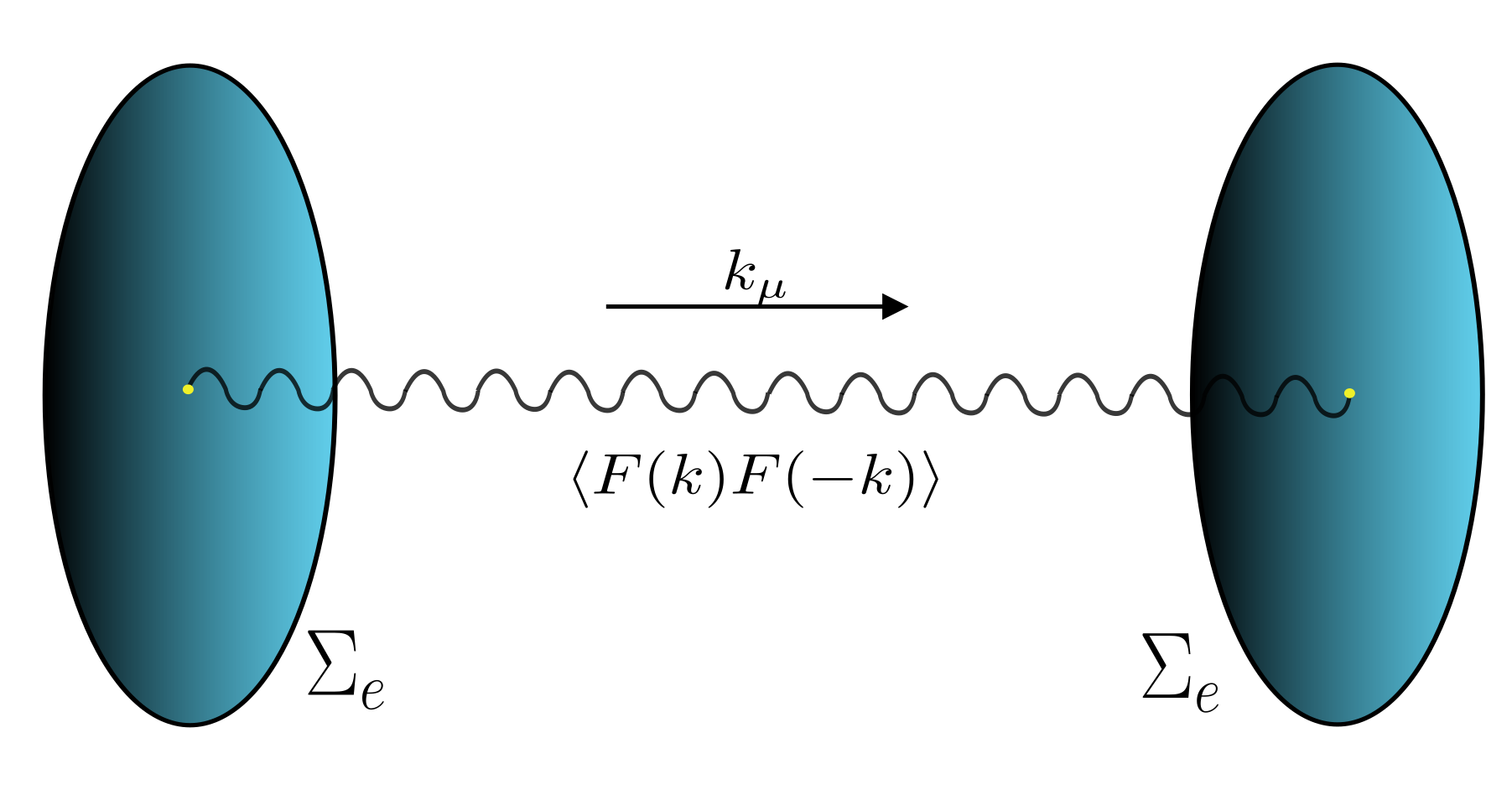}
    \caption{Tree-level amplitude of $\Sigma_e$-$\Sigma_e$ scattering.}
    \label{fig:photon_tree}
\end{figure}

With these Feynman rules, we are in a position to compute the tree-level amplitude for $\Sigma_e$-$\Sigma_e$ scattering. The Feynman diagram is shown in \pref{fig:photon_tree} and we find:
\begin{align}
    iM_{\Sigma_e \Sigma_e} &= V_e^{\mu \nu} \left\langle F_{\mu \nu} F^{\rho \sigma}\right\rangle V^e_{\rho \sigma} \ , \nonumber \\
    & = -2ie^2\Sigma_{e}^{\mu \nu} \frac{k_{\mu}k^{\rho}\delta_{\nu}\,^{\sigma}}{k^2}\Sigma^{e}_{\rho \sigma} \;.
\end{align}
We can rewrite this in terms of the standard electric current as (recall, $k_{\rho}\Sigma_e^{\rho \kappa} = i\, J_e^\kappa$)
\begin{equation}
    iM_{\Sigma_e\Sigma_e} = i M_{ee} = 2ie^2 \frac{J_e \cdot J_e}{k^2} 
\end{equation}
which is precisely the result of \citep{Weinberg:1965rz} upto an overall normalisation due to a difference in conventions.

\begin{figure}[H]
    \centering
    \includegraphics[width=0.5\textwidth]{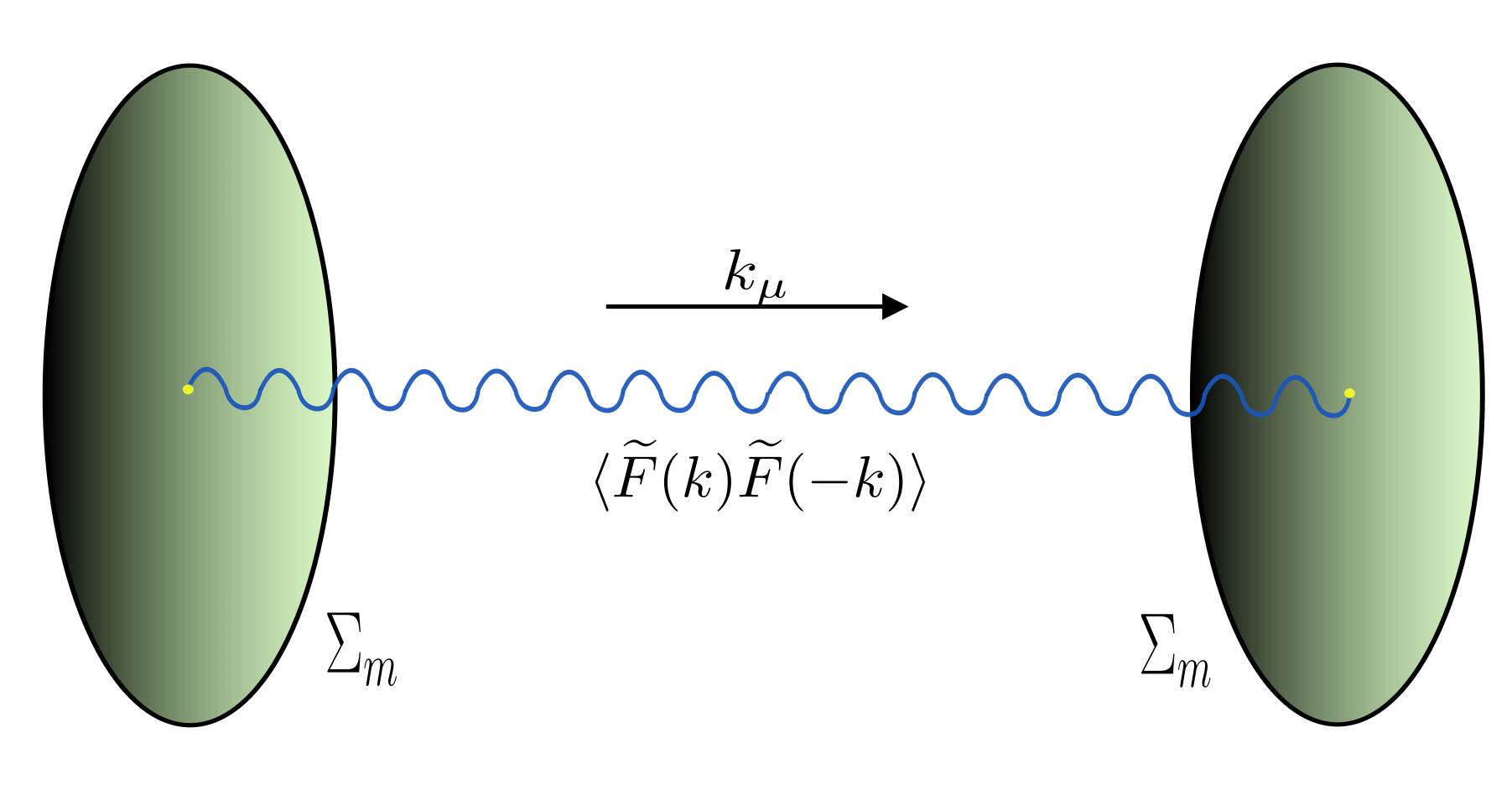}
    \caption{Tree-level amplitude of $\Sigma_m$-$\Sigma_m$ scattering.}
    \label{fig:dual_photon_tree}
\end{figure}

The case of scattering of magnetic sources does not require an independent computation in our approach. Instead, it corresponds to a choice of using $\widetilde{F}_{\mu \nu}$ for photon fields and $\Sigma_{m}$ as its source. By starting from the propagator with $\widetilde{F}$ as photon field and $\Sigma_m$ as its source, one can compute tree-level magnetic-magnetic scattering shown in \pref{fig:dual_photon_tree} to obtain:

\begin{equation}
    iM_{\Sigma_m\Sigma_m} = i M_{gg} = 2ig^2 \frac{J_m \cdot J_m}{k^2} \;.
\end{equation}
This is, naturally, also the result one obtains by duality symmetry from the result of $iM_{ee}$.

Finally, we arrive at the \textit{apparently paradoxical} case of electric-magnetic scattering at tree-level. This amplitude is special in the following sense. As we discussed, in presence of charge quantisation, it is not consistent to carry out a Feynman diagram expansion in both couplings simultaneously. However, for charge quantisation to hold, one needs at least the semi-classical regime where the electromagnetic fields can be treated as classical, but the matter sector is quantum. If both the electromagnetic and the matter sector are classical, which is the case for tree-level amplitudes here, charge quantisation need not be imposed. In other words, classically it is possible to consider the effect of electric and magnetic scattering. Nevertheless, this classical result will not have higher loop counterparts since the  subsequent terms in the amplitudes will go as $\mathcal{O}\Big((eg)^n\Big)$ for all $n >1$ which cannot be captured by perturbation theory due to charge quantisation. 

This can be \textit{prima facie} computed as:

\begin{figure}[H]
    \centering
    \includegraphics[width=0.5\textwidth]{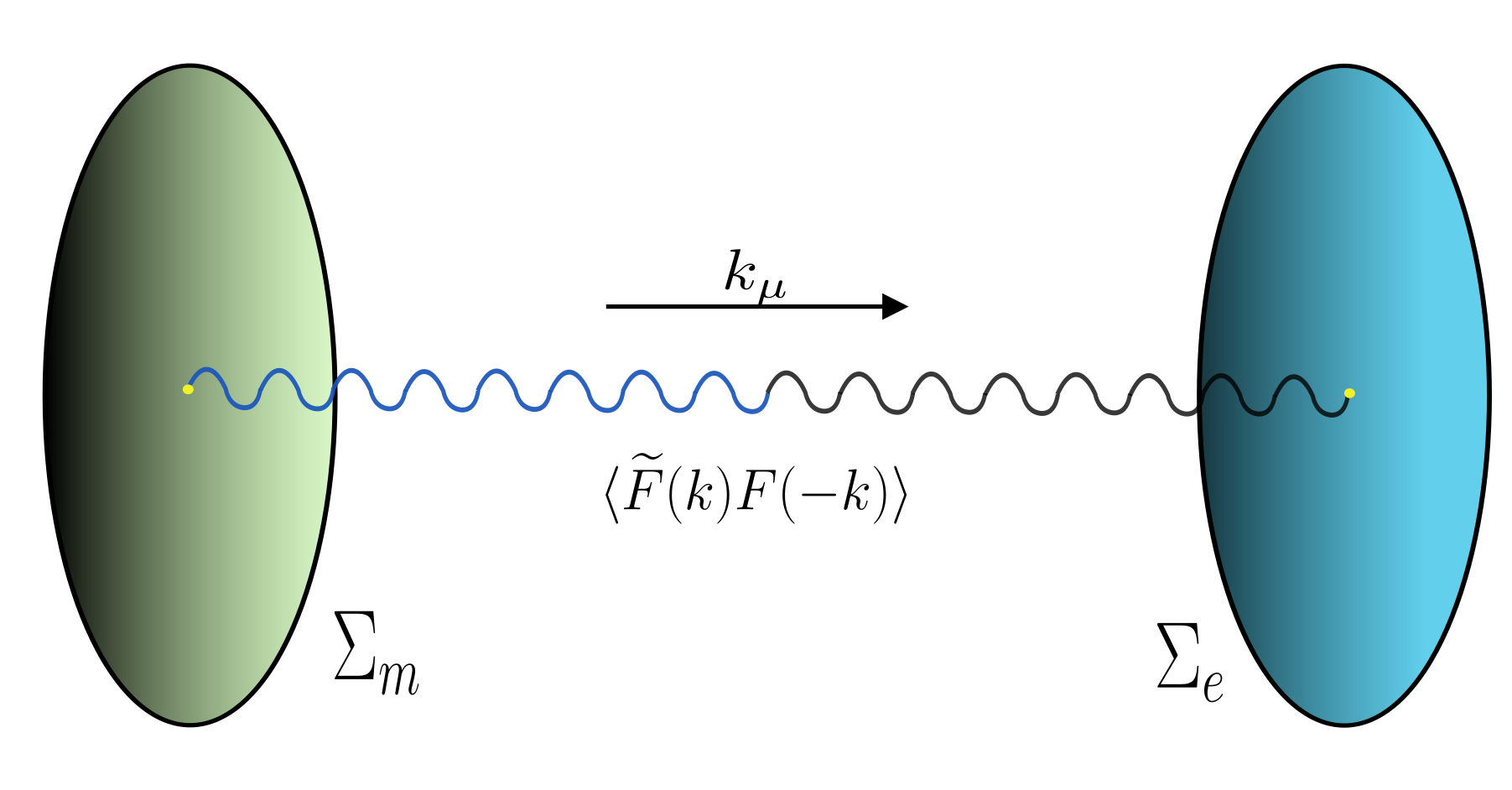}
    \caption{Tree-level amplitude of $\Sigma_m$-$\Sigma_e$ scattering.}
    \label{fig:em_tree}
\end{figure}

\begin{align}
   iM_{\Sigma_e \Sigma_m} = iM_{eg} &=  eg \Sigma_e^{\mu \nu} \langle F_{\mu \nu}\widetilde{F}^{\rho \sigma}\rangle \Sigma^{m}_{\rho \sigma} \nonumber \\
   &= -ieg \frac{(\Sigma_m^{\mu \nu} \epsilon_{\mu\nu \tau \kappa}k ^{\tau}) (k_{\rho}\Sigma_e^{\rho \kappa})}{k^2} \;.
\end{align}

Notice that the answer is perfectly Lorentz- and gauge-invariant, as it should be. However, \emph{if} one insists on re-expressing this answer in terms of the familiar currents that naturally couple to gauge potentials (and which are therefore unnatural from the perspective of our field strength-based approach), we need to do a bit more work. 

As before, we can go ahead and identify $k_{\rho}\Sigma_e^{\rho \kappa} = i\, J_e^\kappa$. But for the other source, we need to invert the identification rules given in \Cref{sec:action-symmetries}. This can be inverted if one makes use of a constant \textit{reference} vector $n_\mu$ and writes
\begin{equation} \label{eq:invert_sigma}
     \Sigma^{\mu \nu}_m=i  \frac{n^\mu J_m^\nu-n^\nu J_m^\mu}{(n \cdot k)} \;.
\end{equation}
The electric-magnetic scattering amplitude then assumes the form
\begin{equation}
    i M_{eg}=  ieg \frac{n^\mu J^{\nu}_m\epsilon_{\mu \nu \rho \sigma} k^\rho J_e^\sigma}{(n\cdot k)} 
\end{equation}
once again in agreement with \citep{Weinberg:1965rz}. But here we also see the resolution of the apparent paradox. The amplitude only appears to be violating Lorentz symmetry (and indeed gauge symmetry) because of our insistence on expressing the answer in terms of the standard currents as sources. Our formalism treats the field strength itself as fundamental, and the sources for the field strength are not the standard currents, but the sources $\Sigma_e$ or $\Sigma_m$, in terms of which the amplitude respects all the symmetries. Additionally, \cref{eq:invert_sigma} is not the unique way of solving for $\Sigma_m$ in terms of $J_m$. In general, the existence of such an inverse is guaranteed in flat spacetime (essentially when Poincar\'{e}'s lemma holds) and is typically expressed in terms of what are called \textit{linear homotopy operators} \cite{Bott:1982xhp}. It is plausible that there might exist alternative ways of solving for $\Sigma$ such that the apparent violation of Lorentz symmetry can be avoided in terms of the usual currents as well. This is an avenue for further investigation that we have not pursued in this paper.

It is noteworthy that \textit{all three amplitudes} computed in this section are, in fact, essentially the same physical quantity, merely expressed in different variables according to one's choice of ``duality frame.'' Furthermore, as we will see in \Cref{sec:charge-renormalisation}, the insistence on using a specific duality frame for a theory that is now formulated in a manifestly duality symmetric manner is at the root of many of the so-called paradoxes in the literature and prior attempts at their resolution. Once a manifestly symmetric action is adopted, as in this paper, all these disagreements disappear --- or, rather, are trivialised --- and we recover consistent, meaningful results following time-tested methods of perturbative quantum field theory.

We now move on to looking at loop computations and specifically the issue of charge renormalisation using the action presented in this paper.

\section{Charge Renormalisation}
\label{sec:charge-renormalisation}
The question of how electric and magnetic charges are renormalised in QEMD has been mired in confusion and debate for decades \citep{Schwinger:1966zza,Schwinger:1966zzb,Coleman:1982cx}. Recently, an attempt to resolve this tension was put forward in \citep{Newey:2024gug}. This resolution required that certain topological contributions were removed \emph{by hand} to arrive at a physically sensible answer. This is unsatisfactory, as a consistent quantum field theory, expanded in a perturbative coupling, should unambiguously lead to an answer for charge renormalisation following textbook methods. Indeed, as we will show in this section, working with the action introduced in \Cref{sec:action-symmetries} and the Feynman rules derived in \Cref{sec:feynman-rules} will lead us to a clear, consistent answer for charge renormalisation in theories with magnetic monopoles. In turn, the manifest duality
symmetry of our action allows us to identify that much of the existing confusion in literature
is due to our adherence to using the terms electric and magnetic in a duality invariant theory and using formalisms that privilege one over the other. Such a distinction is moot in a manifestly duality-invariant theory.

At the level of the classical theory, i.e.,~at tree-level, one need not impose charge quantisation. However, that is no longer true once we start computing the quantum corrections. The charge quantisation condition needs to be imposed at the level of the bare couplings for consistency \citep{Schwinger:1966zza}. In our approach, this necessitates identifying which of the sources $\Sigma_e$ or $\Sigma_m$ is perturbatively coupled. In light of charge quantisation, only one of them can perturbatively couple to the photon (which would then be correspondingly captured by the field strength $F$ or $\widetilde{F}$) and that is the coupling that one can consistently renormalise following standard techniques of perturbative quantum field theory.

\begin{figure}[H]
    \centering
    \includegraphics[width=0.75\textwidth]{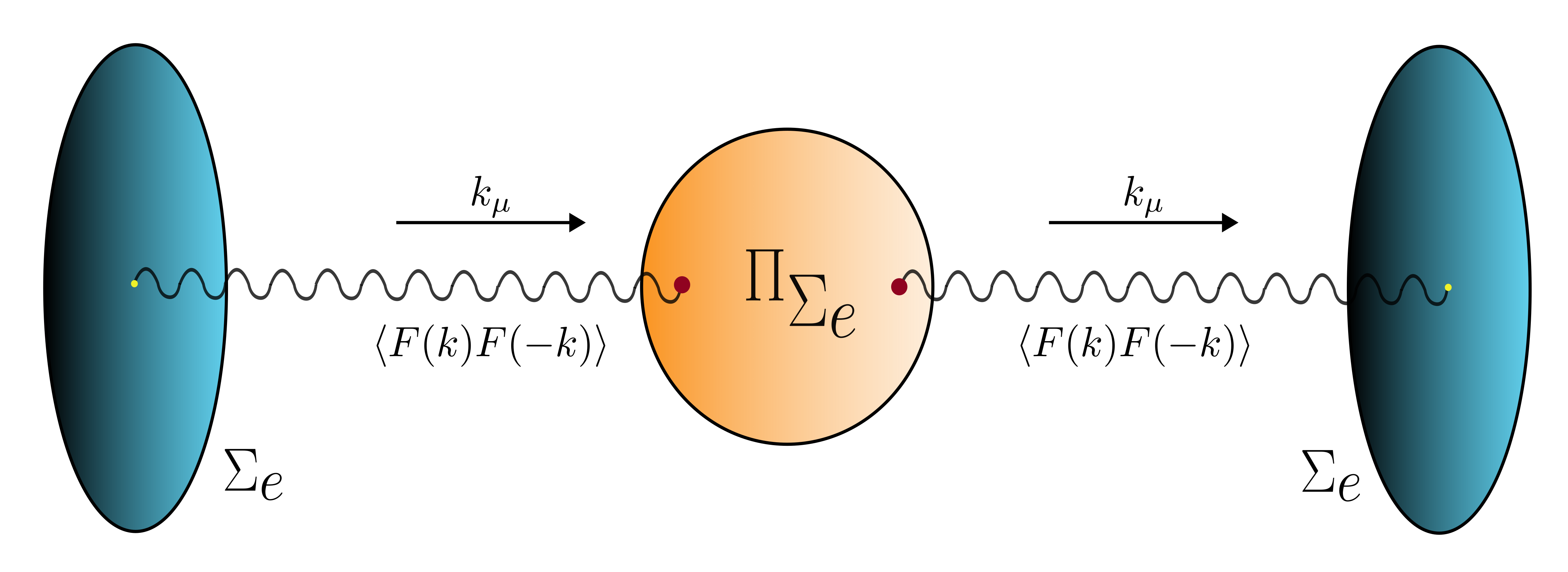}
    \caption{The $1$-loop correction to $\Sigma_e-\Sigma_e$ scattering amplitude.}
    \label{fig:1-loop}
\end{figure}

Let us consider the case when $\Sigma_e$ is perturbatively coupled. Computing the $1$-loop correction to the $\Sigma_e$-$\Sigma_e$ scattering would necessarily require us to supplement our action with a kinetic term for the source. For our purposes, it will suffice to assume that a kinetic term has been provided that is consistent with all the symmetries, and we capture the effect of the loop correction to the photon line by a vacuum polarisation function $\Pi^{[\mu_1 \nu_1] \,;\, [\mu_2 \nu_2]}_{\Sigma_e}(k^2)$ as shown in \pref{fig:1-loop}.

We do not need to know the exact details of the vacuum polarisation function, but the Ward identities for the sources allow us to conclude that the tensor structure is captured entirely by the incoming momenta $k$ and the metric, leaving us with an overall undetermined scalar function of the momentum \citep{Newey:2024gug}. Specifically, with a convenient choice of normalization, 
\begin{equation}
    \Pi^{[\mu_1 \nu_1] \,;\, [\mu_2 \nu_2]}_{\Sigma_e}(k^2) = \frac{1}{2} \frac{k^{[\mu_1} \eta^{\nu1] [\mu_2} k^{\nu_2]}}{k^2} \Pi (k^2) \;.
\end{equation}
It is now straightforward to use the Feynman rules to obtain
\begin{align}
    i M_{\Sigma_e\Sigma_e}^{1\text{-loop}} &= - 4 i e^2 \Sigma^e_{\mu \nu} \langle F^{\mu \nu} F_{\mu_1 \nu_1}\rangle \Pi^{[\mu_1 \nu_1] \,;\, [\mu_2 \nu_2]}_{\Sigma^e}(k^2) \langle F_{\mu_2 \nu_2} F^{\rho \sigma} \rangle \Sigma^e_{\rho \sigma} \ , \nonumber\\
    &=  -2ie^2  \frac{\Sigma^e_{\mu \nu} k^{\nu} k_{\sigma}\Sigma^e_{\sigma \nu}}{k^2} \Pi(k^2) \nonumber \ , \\
    &= 2ie^2\frac{J_e \cdot J_e}{k^2}\Pi(k^2) \ .
\end{align}
Taking into account the tree-level contribution as well, we have up to $1$-loop
\begin{equation} \label{eq:1-loop_full}
   i M^{\text{tree}}_{\Sigma_e\Sigma_e} +  i M^{1\text{-loop}}_{\Sigma_e\Sigma_e} = 2i\bigg(  1 +  \Pi(k^2)\bigg) e^2\frac{J_e \cdot J_e}{k^2} \ .
\end{equation}
This suggests that up to 1-loop, the renormalized coupling $e_R$ is 
\begin{equation} \label{eq:elec_ren}
    e_{_R}^2 = \bigg(  1 +  \Pi(k^2)\bigg) 
 \,e^2 = Z \,e^2 \;.
\end{equation}

Now, if it was $\Sigma_m$ instead that perturbatively coupled to the photon, then an essentially identical calculation due to the duality symmetry would yield
\begin{equation} \label{eq:mag_ren}
    g_{_R}^2 = \bigg(  1 +  \Pi(k^2)\bigg) 
 \,g^2 = Z \,g^2 \;,
\end{equation}
Of course, without making use of the duality symmetry, one can also derive \cref{eq:mag_ren} using the Feynman rules derived in the previous section, and one finds exactly the same answer.

\begin{figure}[htbp]
    \centering
    \includegraphics[width=0.5\textwidth]{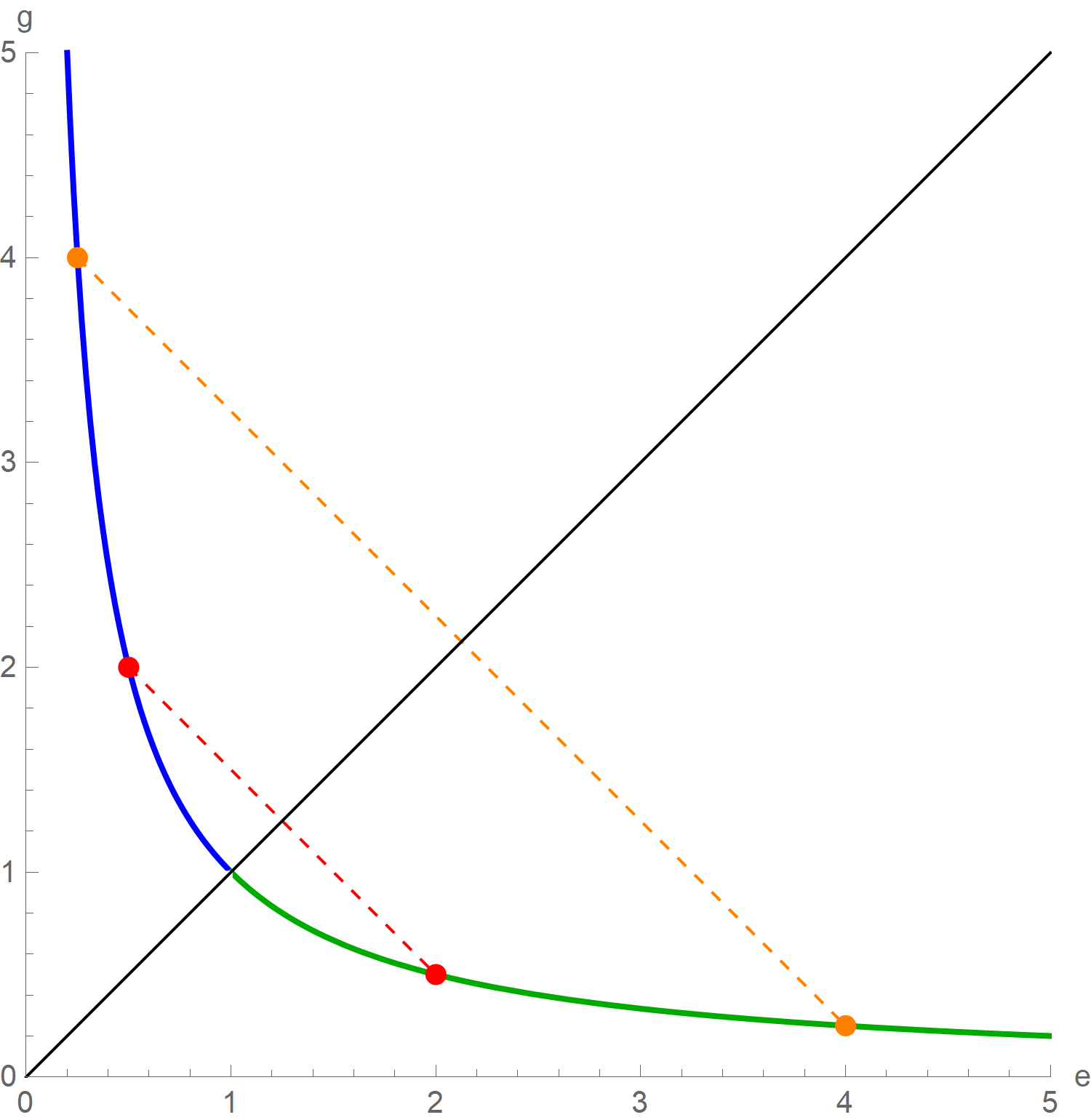}
    \caption{Insisting on using electric and magnetic coupling leads one to a double cover of the actual space of couplings. In this graph, we are working with charges normalised in such a way that the charge quantisation condition reads $eg=1$.}
    \label{fig:em_coupling}
\end{figure}

\begin{figure}[htbp]
    \centering
    \includegraphics[width=0.75\textwidth]{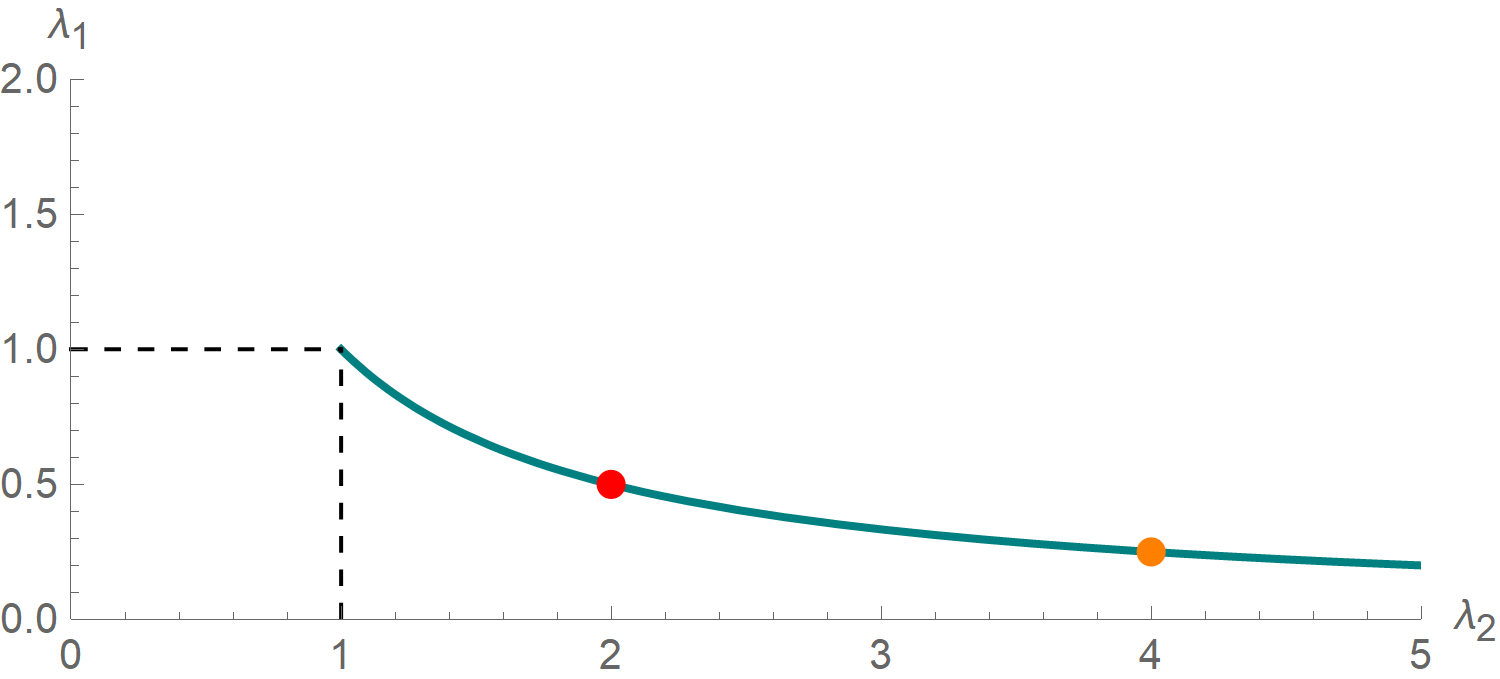}
    \caption{The theory only cares about the distinction between the weak coupling, which we denote by $\lambda_1$, and the strong coupling $\lambda_2$. Charge quantisation (here normalised to $\lambda_1 \lambda_2 = 1$) then further leads us to a single line of possible space of couplings.}
    \label{fig:single_coupling}
\end{figure}

At this point, one might naively conclude that we have recovered Schwinger's result \cite{Schwinger:1966zza,Schwinger:1966zzb} that electric and magnetic charges get renormalised the same way. This conclusion is, as we will argue, hasty and definitely wrong. Indeed, the continued reliance on the terms \textit{electric} and \textit{magnetic} in a duality symmetric theory is at best misleading. If we take the duality symmetric action seriously, then it is simply telling us the following two facts:
\begin{enumerate}
    \item There are two couplings to the photon present in the theory. There is no invariant physical meaning one can associate with the choice to call either coupling \textit{electric} or \textit{magnetic}.
    \item The two couplings at the quantum level are not independent due to charge quantisation; instead, they are controlled by a single perturbative parameter, which one can either think of as the weak coupling or the reciprocal of the strong coupling.
\end{enumerate}
In other words, the continued usage of the terms \textit{electric} and \textit{magnetic} leads one to think of the space of the coupling as in \pref{fig:em_coupling}. The blue curve is the zone where the electric charge is weakly coupled, and the green curve is the region where the magnetic charge is weakly coupled. However, these two regions are not distinct, as they are related by duality. So the blue and the green curves need to be identified as reflected along the line $e=m$ (so the red points will be identified, the orange points will be identified, and so on). This leads to a picture of the actual space of couplings that looks like \pref{fig:single_coupling}, where we denote the weak coupling as $\lambda_1$ and the strong coupling as $\lambda_2$. It does not matter what name we give to each coupling.


In light of this discussion, it should now be clear that the correct interpretation of \pref{eq:elec_ren} and \pref{eq:mag_ren} is that is the coupling associated to the weakly coupled sources (never mind whether it is electric or magnetic) that gets renormalised as 
\begin{equation} \label{eq:weak}
    \lambda_1^{R} = Z \, \lambda_1 \;.
\end{equation}
Depending on one's choice of ``duality frame,'' one may interpret this as the renormalisation of the electric or the magnetic coupling, but \textit{not both simultaneously}.

Furthermore, since the bare couplings must satisfy charge quantisation, i.e.~$\lambda_1 \lambda_2 = 1$, we can just as easily choose to represent our weak coupling parameter as $1/\lambda_2$ instead of $\lambda_1$. Then we can simply rewrite \cref{eq:weak} as
\begin{equation} \label{eq:strong}
    \begin{aligned}
        \frac1{\lambda_2^R} &= Z \, \frac1{\lambda_2} \ , \nonumber \\
      \implies  \lambda_2^R &= Z^{-1} \lambda_2 \;,
    \end{aligned}
 \end{equation}
where one should think of $Z^{-1}$ as being evaluated order by order in perturbation theory. It then follows, rather trivially, that, 
 \begin{equation}
     \lambda_1^R \lambda_2^R = \lambda_1 \lambda_2
 \end{equation}
 or in terms of electric and magnetic couplings,
 \begin{equation}\label{eq:coleman}
     e_{_R} \, g_{_R} = e  \, g
 \end{equation}
which is the result of Coleman \citep{Coleman:1982cx}.

Notice how the renormalisation group invariance of the charge quantisation rule follows rather trivially in this manifestly duality-symmetric formalism. This is not an accident. It is a commonly appreciated fact, often expressed in the adage that duality symmetries only present themselves as non-trivial when they are not manifest or, said differently (and almost tautologically), duality symmetry is trivial in a manifestly duality invariant theory.

We conclude this section with a few general remarks. Despite reaching the same conclusion as \citep{Newey:2024gug}, our derivation differs in several significant ways. Firstly, we do not need to manually remove any terms that contribute to the observable under investigation to arrive at the final result. Our presentation, in contrast, starts with an action and proceeds using the standard rules of perturbative quantum field theory. The difference is most stark in the absence of contact terms in our computation (unlike in \citep{Newey:2024gug}), which is what a consistent computation should give. Secondly, we do not expand simultaneously in both couplings. In the specific action considered here (electromagnetic theory of a fundamental monopole and a fundamental electric charge), this is simply forbidden in our formalism on grounds of consistency. There is only one perturbative coupling, and any Feynman diagram-based computation must expand in the weak coupling. Nevertheless, the manifest duality symmetry allows us to cleanly derive the RG invariance of the charge quantisation rule. Of course, as pointed out in \citep{Newey:2024gug}, there are dark photon models where there are essentially two $\mathrm{U}(1)$ gauge sectors, and one can have kinetically mixed regimes where one can treat both electric and magnetic charges as perturbatively coupled. We are certain that our results can be extended to describe such models as well; that is not the focus of the present article. As it stands, for the simple extension of QED with fundamental monopoles, there are no regions of the coupling space where both electric and magnetic charges weakly couple to the photon. Finally, while we presented explicit results only up to $1$-loop, it is straightforward to see that the basic structure of the RG flow of the couplings remains the same and only $Z$ gets determined to higher and higher orders in perturbation theory. Therefore, indeed the RG invariance of the charge quantisation relation is an \textit{all-loop} result.

It is possible to wonder that turning on the dynamics for sources should necessarily require us to express $\Sigma_e$ in terms of the familiar currents $J_e$. This would seemingly introduce an apparent non-locality (by either using \pref{eq:invert_sigma} or something equivalent) in the action in the coupling terms $\sim F^{\mu \nu} \Sigma^e_{\mu \nu}$. However, the theory is still local which we wish to clarify based on our analysis of this (and the previous) section. Firstly, locality of QFT is decided by the pole structure of its scattering amplitudes and not by the action. Secondly, while the source for the field-strengths are expressed non-locally in terms of the familiar matter currents, one needs to remember they directly couple to the field-strength, as opposed to a gauge field. The virtue of adopting a framework where the field-strength, instead of the gauge field,  is the dynamical variable leads us to Feynman rules which ensures (see \pref{eq:1-loop_full}) (i) the scattering amplitude only has the expected pole structure and (ii) only combinations of $k^\mu\Sigma^e_{\mu \nu} \sim J_\nu^e$ appears in the scattering amplitude which is perfectly local. The resulting S-matrix is that of a local quantum field theory and expressed only in terms of the familiar currents, as it should be.

\section{Conclusion}

The central result of this article is to point out a manifestly duality-invariant, Lorentz invariant, and local action for describing QEMD, the theory of light interacting with electric and magnetic charges. Such an action treats field strengths, as opposed to gauge potentials, as fundamental variables, something that the requirement of duality invariance necessitates. We explicitly show that once such an action is adopted, standard perturbative quantum field theory techniques yield consistent answers at tree- and loop-level without any additional nuances or inputs. This is in stark contrast with \textit{all} prior attempts to describe QEMD. 

While we focused only on flat four-dimensional Minkowski spacetime in this paper, we want to highlight two important generalisations. First, the action we have discussed can easily be coupled to curved spacetime. Indeed, the action obtained from dimensional reduction from $6$D on a torus in \citep{Sen:2019qit} was written in a general curved $4$D Lorentzian spacetime. Sen-type theories have a slightly unusual way of coupling to curved spacetime so as to ensure the additional fields do not gravitate. This is very well understood by now, \citep{Sen:2015nph,Sen:2019qit,Andriolo:2020ykk,Lambert:2019diy,Lambert:2023qgs,Chakrabarti:2022jcb,Chakrabarti:2020dhv,Chakrabarti:2023czz,Gustavsson:2020ugb,Andrianopoli:2022bzr,Vanichchapongjaroen:2020wza,Phonchantuek:2023iao} and indeed we have checked that there is a way to couple the action considered here to dynamical spacetime \textit{without} needing to dimensionally reduce a $6$D theory. We may revisit this question later in a separate article. Furthermore, while we only derived the decoupling of the additional fields from the physical fields in momentum space within the purview of perturbation theory, this decoupling of the extra fields is a \textit{non-perturbative} result that can be best seen with a Hamiltonian analysis. The analysis is identical to the case of self-dual forms as shown in \citep{Sen:2019qit} and we refer interested readers to it.

Secondly, it is known that in string theory, magnetically charged D-branes inevitably exist with their electrically charged cousins. Indeed, both are charged under the same abelian Ramond-Ramond field strengths, and their charges follow a quantisation rule as well. Recently these topics were revisited within the context of Sen's formalism in \citep{Hull:2025yww}. We want to point out that the action considered here can be straightforwardly generalized to accommodate such higher-dimensional instances. This is best seen if we rewrite the action in the differential form notation, where in any dimension $D$ it assumes the form
\begin{equation}
    S = \frac1{2} \sum_{i=1}^2 \int G^{(i)} \wedge \star G^{(i)} - \int G^{(1)} \wedge \star F - \int G^{(2)} \wedge \star \widetilde{F} -  \int \widetilde{F} \wedge \star \Sigma - \int F \wedge \star \widetilde{\Sigma} \;.
\end{equation}
Now, we have $G^{(i)} = \d B^{(i)}$ to be a pair of $(p+1)$-form field-strengths, $F$ is also a $(p+1)$-form field strength-like variable with its dual $\star F := \widetilde{F}$ being a $(d-p-1)$-form. Naturally the source $\Sigma$ is now a $(d-p-1)$-form source due to a magnetically charged $(d-p-3)$-brane and the dual source $\star \Sigma := \widetilde{\Sigma}$ is a $(p+1)$-form due to an electrically charged $(p-1)$-brane. In string theory only certain values of $p$ are allowed once we choose $D=10$ and decide on the chirality of supercharges. It would be interesting to see if there are any other constraints that could already be imposed from anomalies (or lack thereof) of these extended objects with respect to various gauge and higher-form symmetries they will possess in other dimensions without imposing supersymmetry. 

It has recently been demonstrated that the charge quantisation rule is intimately related to the notion of generalised symmetries \citep{Hull:2024uwz} (also see \citep{Lechner:1999ga}).\footnote{We would like to thank C.~Hull for bringing reference \citep{Hull:2024uwz} to our attention.} It would be interesting to understand the role of generalised symmetry and its interplay with charge quantisation for the framework presented in this paper. We hope to return to this question in future.

Several phenomenological consequences of the existence of magnetic monopoles have often been hard to pin down (e.g.~the Callan-Rubakov effect \citep{Callan:1982ac,Callan:1982ah,Callan:1982au,Rubakov:1982fp,Rubakov:1988aq}) that could be attributed to the lack of a fully satisfactory action principle. We expect that our result would be immensely helpful in providing clear, uncontroversial predictions for such cases. Incidentally, while we specifically focused on the case of a fundamental monopole, our action should be equally adept as an effective field theory of a composite (i.e.~'t Hooft-Polyakov) monopole as long as one is probing a distance scale much larger than the size of the monopole core. These often include exciting, dynamically rich theories, such as Seiberg-Witten theories, where, once again, in a suitable regime, we expect our formalism to be very useful. Especially, electric-magnetic duality is the simplest example of strong-weak duality (S-duality), which plays a pivotal role in our understanding of non-perturbative aspects of superstring theories, as well as several supersymmetric gauge theories. In the presence of such dualities, it is known that one can improve any perturbation series by expanding around both the weak and the strong coupling points and use resurgence techniques to obtain a much better approximation than either of these perturbation series would provide individually \citep{Sen:2013oza,Beem:2013hha,Honda:2014bza,Honda:2015ewa}. Given our formalism, it would be instructive to carry out a similar analysis and extract non-perturbative answers for QEMD. We hope to return to some of these questions in future works.

\section*{Acknowledgements}
We thank Klaus Bering for discussions and helpful feedback to an earlier version of the manuscript. AA and SC acknowledge the financial support by the Czech Science Foundation (GA\v{C}R) grant “Dualities and higher derivatives” (GA23-06498S). MR is supported by an Inspire Faculty Fellowship. 

\appendix

\section{Maxwell's Equations With $2$-form Sources} \label{app:Sen}

In flat spacetime, due to Poincar\'e lemma, any conserved current $1$-form can be expressed in terms of a divergence of a $2$-form. Essentially,

\begin{equation}
    \partial_\mu J^\mu = 0 \implies J^\mu = \partial_\nu \Sigma^{\mu \nu} \;.
\end{equation}
Notice this $2$-form source $\Sigma$ (also called a Stokes' surface) is not unique. It has a ``gauge'' symmetry
\begin{equation}
    \Sigma^{\mu \nu} \to \Sigma'^{\mu \nu} = \Sigma^{\mu \nu} + \partial_\rho \Gamma^{\rho \mu \nu}
\end{equation}
with a $3$-form gauge parameter $\Gamma^{\rho \mu \nu}$ that leaves the conservation equation for $J$ unchanged. Therefore, physically, $\Sigma$ and $\Sigma'$ are equivalent.

Now the extended Maxwell's equations read
\begin{align}
    \partial_\mu F^{\mu \nu} &= e J_e^\nu \\
    \partial_\mu \widetilde{F}^{\mu \nu} &= g J_m^\nu \;.
\end{align}
We can rewrite this as 
\begin{align}
    \partial_\mu (F^{\mu \nu} - e \Sigma_e^{\mu\nu}) &= 0 \\
    \partial_\mu (\widetilde{F}^{\mu \nu} - g \Sigma_m^{\mu\nu}) &=0 \;.
\end{align}
Once again by Poincar\'e lemma this implies 
\begin{align}
    F^{\mu \nu} - e \Sigma_e^{\mu\nu} &= \partial_\rho \Gamma_e^{\rho \mu \nu} \implies F^{\mu \nu} = e \Sigma_e^{\mu\nu} + \partial_\rho \Gamma_e^{\rho \mu \nu} \\
    \widetilde{F}^{\mu \nu} - g \Sigma_m^{\mu\nu} &= \partial_\rho \Gamma_m^{\rho \mu \nu} \implies \widetilde{F}^{\mu \nu} = g \Sigma_m^{\mu\nu} + \partial_\rho \Gamma_m^{\rho \mu \nu} \;.
\end{align}
Taking Hodge dual of the latter and using the former, we see
\begin{equation}
    e \Sigma_e^{\mu \nu} = - g \widetilde{\Sigma}_m - \partial_\rho\Big(\frac{1}{2}\epsilon^{\mu \nu}_{\;\;\;\kappa \sigma} \Gamma_m^{\rho \kappa \sigma} + \Gamma_e^{\rho \mu \nu}     \Big) = - g \widetilde{\Sigma}_m + \partial_\rho \widehat{\Gamma}^{\rho \mu \nu} \;.
\end{equation}
Therefore, the electric and magnetic Stokes' surfaces are related by Hodge duality (up to the gauge equivalence) and it follows directly from Maxwell's equation. Notice this derivation do not assume anything about the original currents $J_e$ and $J_m$ except that they are separately conserved.
\bibliography{refs}
\end{document}